# Microgrid Optimal Energy Scheduling Considering Neural Network based Battery Degradation

Cunzhi Zhao, *Student Member, IEEE* and Xingpeng Li, *Senior Member, IEEE*

*Abstract*—Battery energy storage system (BESS) can effectively mitigate the uncertainty of variable renewable generation. Degradation is unpreventable and hard to model and predict for batteries such as the most popular Lithium-ion battery (LiB). In this paper, we propose a data driven method to predict the battery degradation per a given scheduled battery operational profile. Particularly, a neural network based battery degradation (NNBD) model is proposed to quantify the battery degradation with inputs of major battery degradation factors. When incorporating the proposed NNBD model into microgrid day-ahead scheduling (MDS), we can establish a battery degradation based MDS (BDMDS) model that can consider the equivalent battery degradation cost precisely with the proposed cycle based battery usage processing (CBUP) method for the NNBD model. Since the proposed NNBD model is highly non-linear and non-convex, BDMDS would be very hard to solve. To address this issue, a neural network and optimization decoupled heuristic (NNODH) algorithm is proposed in this paper to effectively solve this neural network embedded optimization problem. Simulation results demonstrate that the proposed NNODH algorithm is able to obtain the optimal solution with lowest total cost including normal operation cost and battery degradation cost.

*Index Terms*— Battery degradation, Battery energy storage system, Energy management system, Machine learning, Microgrid day-ahead scheduling, Neural network, Optimization.

## NOMENCLATURE

*Sets:*
$S_T$　　Set of time intervals.
$S_G$　　Set of controllable micro generators.
$S_S$　　Set of energy storage systems.
$S_{WT}$　Set of wind turbines.
$S_{PV}$　Set of PV systems.

*Parameters:*
$c_{Gi}$　　Linear cost for controllable unit *i*.
$c_{Gi}^{NL}$　No load cost for controllable unit *i*.
$c_{Gi}^{SU}$　Start-up cost for controllable unit *i*.
$\Delta T$　　Length of a single dispatch interval.
$R_{prcnt}$　Percentage of the backup power to the total power.
$E_{Si}^{Max}$　Maximum energy capacity of ESS *i*.
$E_{Si}^{min}$　Minimum energy capacity of ESS *i*.
$c_{Buy}^t$　Wholesale electricity purchase price in time interval *t*.
$c_{Sell}^t$　Wholesale electricity sell price in time interval *t*.
$P_{Gi}^{Max}$　Maximum capacity of generator *i*.
$P_{Gi}^{Min}$　Minimum capacity of generator *i*.
$P_{Grid}^{Max}$　Maximum thermal limit of tie-line between main grid and microgrid.
$P_{Gi}^{Ramp}$　Ramping limit of diesel generator *i*.
$P_{Si}^{Max}$　Maximum charge/discharge power of BESS *i*.
$P_{Si}^{Min}$　Minimum charge/discharge power of BESS *i*.
$\eta_{Si}^{Disc}$　Discharge efficiency of BESS *i*.
$\eta_{Si}^{Char}$　Minimum capacity of generator *i*.

*Variables:*
$U_{Buy}^t$　Status of buying power from main grid in time interval *t*.
$U_{Sell}^t$　Status of selling power to main grid status in time *t*.
$U_{Char}^{t,i}$　Charging status of energy storage system i in time interval t. It is 1 if charging status; otherwise 0.
$U_{Disc}^{t,i}$　Discharging status of energy storage system i in time interval t. It is 1 if discharging status; otherwise 0.
$U_G^{ti}$　Status of generator *i* in time interval *t*. It is 1 if on status; otherwise 0.
$V_G^{ti}$　Startup indicator of Status of generator *i* in time interval *t*. It is 1 if unit i starts up; otherwise 0.
$P_{Gi}^t$　Output of generator *i* in time interval *t*.
$P_{Buy}^t$　Amount of power purchased from main grid power in time interval t.
$P_{Sell}^t$　Amount of power sold to main grid power in time interval t.
$P_L^t$　Demand in the microgrid in time interval *t*.
$P_{Disc}^{t,i}$　Discharging power of energy storage system *i* at time *t*.
$P_{Char}^{t,i}$　Charging power of energy storage system i at time t.

*Abbreviations*:
RES　Renewable energy sources.
BESS　Battery energy storage system.
LiB　Li-ion battery.
SOC　State of charge.
SOH　State of health.
DOD　Depth of discharge.
EV　Electric Vehicle.
MDS　Microgrid day-ahead scheduling.
NNBD　Neural network based battery degradation.
BDMDS　Battery degradation based MDS.
CBUP　Cycle based battery usage processing.
NNODH　Neural network and optimization decoupled heuristic.
CMDS　Conserved MDS.

## I. INTRODUCTION

Renewable energy sources (RES) will play an important role in future microgrid due to the 3-D (decentralization, decarbonation, and digitalization) trend. The proportion of RES in power generation is growing significantly. However, the system stability is substantially weakened by the stochastic

Cunzhi Zhao and Xingpeng Li are with the Department of Electrical and Computer Engineering, University of Houston, Houston, TX, 77204, USA (e-mail: czhao20@uh.edu; Xingpeng.Li@asu.edu).

and intermittent generation of high penetration RES. Battery energy storage system (BESS) is an effective flexible solution for addressing the uncertainty of variable RES induced system [1]. Thus, much more BESSs will be available in bulk power systems and small-scale microgrids in near future.

The characteristic of rechargeable chemical battery makes it degrade during cycling. This degradation can be accelerated by extreme fast charging or discharging cycles, extreme low or high ambient temperature and over charging or over discharging. However, the internal states of battery remain difficult to estimate while the battery is taking a more important role as BESS in power energy systems [2]. Therefore, the battery degradation is quite hard to predict. The working principle of a BESS is similar to a voltage source in series with impedances. However, the operating conditions and environments are various for BESS in microgrids, as well as in bulk power systems. Thus, the BESS model cannot be simply treated as a voltage-source based model [3]. Li-ion battery (LiB) has been widely used as energy storage due to its high energy density and low memory effect nature. LiB degrades mainly because of the loss of Li-ions, the loss of electrolyte and the increase of internal resistance [4]. During battery cycling, the influential factors that cause the degradation include the battery operating ambient temperature, charging/discharging rate, state of charge (SOC), state of health (SOH), and depth of discharge (DOD) [5]-[6].

Previous studies have developed some energy management strategies for BESS-integrated microgrids. In [7]-[14], it has been proved that the BESS can be seamlessly integrated into microgrid especially for those microgrids with high penetration renewable energy sources. However, battery degradation is not considered in those energy management strategies.

Some simple models of battery degradation are proposed in the literature. In [15], DOD is used to calculate the remaining useful cycles. The battery degradation is estimated based on the remaining useful battery cycle and the actual capacity. The only variable considered in [15] is the DOD and they assume the degradation process to be linear throughout the battery life which is not reasonable. Similar to [15], [16]-[19] proposed some DOD based models to estimate the battery degradation for each cycle which omits other important degradation factors. Pascali in [20] adopts the Butler-Volmer equation for battery degradation model to illustrate the diffusion of the solvent reactants. However, the degradation data are based on the experiment for different discharge currents and SOC values, which are not sufficient as they omit other important factors contributing to battery degradation. As mentioned in [21]-[24], the linear assumption of battery degradation may simplify the problem and reduce the computational difficulty, but the degradation value may not be accurately predicted. In summary, these popular heuristic battery degradation models can be represented as two battery degradation models: (i) linear degradation model that applies a linear degradation cost based on the power output or the energy usage which may lead to a large error on battery degradation quantification, and (ii) DOD and/or SOC based model that considers DOD and/or SOC as the input which omits the other important degradation factors such as charge/discharge rate and ambient temperature.

A data driven, comprehensive battery degradation model is developed in [25]. Although the accuracy of the prediction for battery degradation is high, the developed model is hard to incorporate with the various operation of BESS due to the limited battery degradation data. A temperature based battery degradation model is proposed in [26] for electric vehicle (EV). However, the model is only reflected by the internal temperature of EV's battery pack and may not suitable for the BESS. Saldana's paper [27] proposed a battery degradation matrix reference for EV. However, it is impractical to consider this model in the microgrid day-ahead scheduling (MDS) problem due to the computational complexity in the optimization problem. The electric vehicle has been researched and conducted in the V2G system in [28]-[30] due to the similarity between EV and BESS. However, an important EV battery degradation factor, the charge/discharge rate, is neglected in those papers which may lead to an inaccurate battery degradation prediction. A data driven degradation model is presented in [31]. A quadratic equation is formed based on the collected data; however, the data are collected only under different profiles of DOD and SOC which omits other degradation related factors. References [32]-[34] present some advanced methodology to predict the lifetime of the battery cell and then the degradation can be averaged for each cycle. However, those methodologies are specially designed for the battery aging tests in which the battery's cycle is set as a fixed charge or discharge rate. In other words, the battery degradation or remaining cycle prediction is based on the fixed charging or discharging cycles in those methodologies. Therefore, they cannot reflect the degradation prediction for usage-based BESS due to the battery's dynamic schedule such as various charge or discharge rates at different time intervals in power system/microgrid applications.

In summary, there are mainly three gaps for all the aforementioned battery degradation models, which are addressed in this paper:

1) Existing models do not consider all major critical degradation factors. They focus on only a couple of variables and ignore other critical factors, which limits the accuracy of their battery degradation models.
2) BESS operations are often very dynamic; for instance, they may have very different charge/discharge rates and SOC levels at different time intervals. This is not respected by existing methods that are unable to accurately consider such dynamics in usage-based battery degradation prediction.
3) Some existing degradation models are not positioned to be efficiently incorporated into MDS.

To address the above-mentioned gaps, a fully connected neural network (NN) is proposed to train a battery degradation model, and a cycle based battery usage processing (CBUP) method is developed for the BESS scheduling to accurately predict the battery degradation with the proposed NN model. The input of the NN model is a vector of five features including ambient temperature, charging/discharging rate, SOC, DOD and SOH. This neural network based battery degradation (NNBD) model contains non-linear activation functions in the hidden layers, which makes it complex. When incorporating the proposed NNBD model into MDS, we can establish a bat-tery degradation based MDS (BDMDS) model that can consider the equivalent battery degradation cost.

The BESS scheduling in MDS does not always operate in a fixed charging or discharging cycle. Therefore, the CBUP method is designed to fill the research gap between the fixed-cycle based NNBD model and the dynamic BESS operation in MDS. However, such a complex neural network embedded optimization problem would be hard to solve directly. To address this issue, a neural network and optimization decoupled heuristic (NNODH) algorithm is proposed in this paper to effectively solve the BDMDS problem. The proposed NNODH algorithm iteratively solves the transformed BDMDS problem that is decoupled to the battery degradation calculation and MDS optimization problems. BESS operation constraints with tighter bounds will be generated in each iteration to limit the usage of BESS which can reduce the battery degradation and the relevant cost in the next iteration. However, the microgrid's operation cost will increase if the BESS usage is limited. The goal of the proposed NNODH algorithm is to find the lowest value for the sum of battery degradation cost and microgrid operation cost. It can also record the total cost for each iteration and locate the vertex point which is also the optimal solution for the BDMDS problem. Three benchmark models are also developed to test and compare the performance of the proposed NNODH algorithm. The main contributions of this paper are summarized as follows:

- A set of battery cycle generators is designed to simulate battery degradation under different battery operational profiles. The key features (ambient temperature, charge/discharge rate, SOC level, DOD and updated battery energy capacity) that affect battery degradation are collected for each cycle.
- A neural network based battery degradation model with the above five input features is proposed in this paper and it is able to accurately predict the degradation respect to the current maximum battery energy capacity.
- A cycle based battery usage processing method is proposed to process the BESS profile to correctly incorporate the proposed NNBD model into MDS, addressing the inconsistence between the fixed-cycle based NNBD model and BESS scheduling.
- A BDMDS model is proposed to capture the effect of battery degradation by incorporating the proposed NNBD model into microgrid energy management.
- An NNODH algorithm is proposed to efficiently solve the battery degradation based MDS model that is hard to solve directly. Four battery usage-limiting MDS models, referred to as conserved MDS (CMDS), are developed and used by the NNODH algorithm. The optimal scheduling obtained with NNODH leads to the lowest total cost including the battery degradation cost and microgrid operation cost.
- Validation of the performance for the proposed NNODH algorithm is conducted. Case studies prove that by limiting the battery operation that leads to lower degradation, the total cost can be reduced significantly.

The rest of the paper is organized as follows. Section II presents a traditional MDS model. Section III presents the neural network structure of the proposed battery degradation model and Section IV presents the proposed microgrid energy management strategies with the BESS degradation model. Section V discusses the microgrid testbed, simulation results and sensitivity analysis. Section VI concludes the paper.

## II. TRADITIONAL MICROGRID DAY-AHEAD SCHEDULING

A traditional MDS model is established as a basic model to gauge the proposed BDMDS model. This traditional MDS model consists of (1)-(15) as described below and it does not consider battery degradation.

The objective of this traditional MDS model is to minimize the total cost of the microgrid operations as illustrated in (1). The power balance equation involving controllable generators, renewable energy sources, power exchange with the main grid, BESS output and the load is shown in (2). Constraint (3) enforces the power limits of the controllable units such as diesel generators. The ramping up and down limits are enforced by (4) and (5). Equation (6) ensures the status of power exchange between microgrid and main grid to be either purchasing or selling or stay idle. The thermal limit of the tie-line is enforced by constraints (7)-(8). Equation (9) restricts the BESS to be either in charging mode or in discharging mode or stay idle. Constraints (10)-(11) limit the charging/discharging power of BESS. As shown in (12), the SOC level of BESS can be calculated based on current energy stored in BESS. Equation (13) calculates the energy stored in the BESS for each time interval. The ending SOC level of BESS is forced to be equal to the initial SOC value (14). Constraint (15) ensures the microgrid to have sufficient backup power to address outage events.

Objective:

$$f^{MG} = \sum\sum(P_{Gi}^t c_{Gi} + U_{Gi} c_{Gi}^{NL} + V_{Gi} c_{Gi}^{SU}) \\ + P_{Buy}^t c_{Buy}^t - P_{Sell}^t c_{Sell}^t, \forall i, t \quad (1)$$

Constraints are as follows:

$$P_{Buy}^t + \sum_{i \in S_G} P_{Gi}^t + \sum_{i \in S_{WT}} P_{WTi}^t + \sum_{i \in S_{PV}} P_{PVi}^t \\ + \sum_{i \in S_S} P_{Disc}^{t,i} \\ = P_{Sell}^t + \sum_{i \in S_L} P_{Li}^t \\ + \sum_{i \in S_S} P_{Char}^{t,i} \quad (2)$$

$$P_{Gi}^{Min} \leq P_{Gi}^t \leq P_{Gi}^{Max}, \forall i, t \quad (3)$$

$$P_{Gi}^{t+1} - P_{Gi}^t \leq \Delta T \cdot P_{Gi}^{Ramp}, \forall i, t \quad (4)$$

$$P_{Gi}^t - P_{Gi}^{t+1} \leq \Delta T \cdot P_{Gi}^{Ramp}, \forall i, t \quad (5)$$

$$U_{Buy}^t + U_{Sell}^t \leq 1, \forall t \quad (6)$$

$$0 \leq P_{Buy}^t \leq U_{Buy}^t \cdot P_{Grid}^{Max}, \forall t \quad (7)$$

$$0 \leq P_{Sell}^t \leq U_{Sell}^t \cdot P_{Grid}^{Max}, \forall t \quad (8)$$

$$U_{Disc}^{t,i} + U_{Char}^{t,i} \leq 1, \forall i, t \quad (9)$$

$$U_{Char}^{t,i} \cdot P_{Si}^{Min} \leq P_{Char}^{t,i} \leq U_{Char}^{t,i} \cdot P_{Si}^{Max}, \forall i, t \quad (10)$$

$$U_{Disc}^{t,i} \cdot P_{Si}^{Min} \leq P_{Disc}^{t,i} \leq U_{Disc}^{t,i} \cdot P_{Si}^{Max}, \forall i, t \quad (11)$$

$$SOC_{Si}^t = E_{Si}^t / E_{Si}^{Max}, \forall i, t \quad (12)$$

$$E_{Si}^t - E_{Si}^{t-1} + \Delta T \cdot (P_{Disc}^{t-1,i}/\eta_{Si}^{Disc} - P_{Char}^{t-1,i}\eta_{Si}^{Char}) \\ = 0, \forall i, t \quad (13)$$

$$E_{Si}^{t=24} = E_{Si}^{Initial}, \forall i \quad (14)$$

$$P_{Grid}^{Max} - P_{Buy}^t + P_{Sell}^t + \sum_{G \in S_G}(P_{Gi}^{Max} - P_{Gi}^t)$$
$$\geq R_{percent}\left(\sum_{i \in S_L} P_{Li}^t\right), \forall i, t \quad (15)$$

## III. NEURAL NETWORK BASED BATTERY DEGRADATION MODELING

The traditional microgrid day-ahead scheduling model determines the optimal operational profiles for BESS, controllable generators and tie-line exchange power. However, the BESS in traditional MDS model is considered to be ideal without any degradation and the equivalent battery degradation cost is zero. This may substantially accelerate the aging and replacement of expensive BESS which may lead to economic loss in the long run. Besides, the BESS degradation needs to be accurately quantified for various battery daily operational profiles to obtain the truly optimal scheduling solutions. Thus, a deep learning method, particularly a deep neural network, is proposed in this section to accurately predict the BESS degradation.

### A. Input Data

The deep neural network is applied to learn and predict the battery degradation with several critical features. A battery model implemented in the MATLAB Simulink [35] is used to conduct the battery aging tests. The aging test model is modelled with the battery model and cycling generator in Simulink. The SOC variable can be adjusted in the battery setting. The battery model of Simulink contains several default types of batteries and the battery parameters can also be adjusted with the manufacture's datasheet if we want to introduce a new type of battery. It can also simulate the ambient temperature effects and aging effects of battery. A dynamic internal resistance that is highly related to the battery degradation is also simulated within the battery model. This battery model can simulate various types and configurations of batteries, various conditions, and operating profiles. Built upon this battery model, a battery cycle generator is designed to simulate the charging and discharging cycles under preset charging/discharging rates. For each battery aging test, each cycle is simulated to discharge from a certain SOC to a certain lower SOC and then charge back to the starting SOC. The data collected from the battery aging tests include the ambient temperature, charging/discharging rate, SOC, DOD and reduced energy capacity level. The battery's energy capacity level for each cycle is also used to calculate each cycle's battery degradation value. Table I lists the numbers of battery aging tests that are simulated under different SOC and DOD. Each battery aging test represents a group of simulated battery profile under initial SOC and fixed DOD until the battery capacity degrades to 80% of the maximum rated capacity. This indicates each battery aging test may contain different numbers of cycles. It is common to consider the 80% of the rated capacity as the end of battery's life in battery aging test [36]. For instance, the "4" in Table I means there are four battery aging tests that are simulated at an initial SOC level of 100% with a DOD level of 10%. Moreover, each of these SOC&DOD combinations will be simulated under different ambient temperatures and charging/discharging rates. The charging/discharging rate is also referred as C rate. As shown in (16), C rate measures the speed at which a battery is fully charged or discharged.

$$C\ rate = \frac{I_{Battery}}{E_0} \quad (16)$$

where $I_{Battery}$ represents the charging/discharging current of the BESS and $E_0$ denotes the rated capacity of BESS. For example, 2C means the battery will discharge the full capacity in 0.5 hour. This work conducted 945 different battery aging tests with different values of degradation factors.

Table I Numbers of battery aging tests under various SOC and DOD levels

| DOD | Initial SOC | | | | | |
|---|---|---|---|---|---|---|
|  | 100% | 80% | 60% | 50% | 40% | 20% |
| 20% | 4 | 5 | 17 | 23 | 22 | 23 |
| 30% | 20 | 34 | 36 | 32 | 37 | / |
| 40% | 36 | 41 | 41 | 40 | 44 | / |
| 50% | 38 | 41 | 36 | 42 | / | / |
| 60% | 37 | 37 | 37 | / | / | / |
| 70% | 41 | 36 | / | / | / | / |
| 80% | 39 | 35 | / | / | / | / |
| 90% | 35 | / | / | / | / | / |
| 100% | 36 | / | / | / | / | / |

### B. Data Pre-processing

The battery capacity level or SOH in percent at the end of each cycle is recorded from each simulation of battery aging test. Battery degradation per cycle is defined as the difference between the initial SOH and ending SOH for a given cycle. However, for some battery aging tests with low C rates and small DODs, the degradation for some cycles may be too less to measure and could even be zero. Also, some of the degradation data contain outlier points which may require some further process for a better training result. The battery degradation data are processed with different methods: (i) raw, (ii) smoothed method, and (iii) regressed method, as shown in Fig. 1. "Raw" represents the original data without any pre-processing and the associated training results can be used as a benchmark to gauge the effectiveness of the other two proposed data pre-processing methods. The smoothed method filters out the outliers of the raw data. Regressed method applies linear regression on the smoothed data. All three groups of input data are normalized to increase the training efficiency. The normalization used in [37] is applied in the data pre-processing and shown in (17):

$$\widehat{x_k} = \frac{x_k - E(x_k)}{\sqrt{Var(x_k)}} \quad (17)$$

where $E(x_k)$ represents the expectation of $x_k$ and $Var(x_k)$ represents the variance of the $x_k$. The input data will be split into two parts, 80% as the training dataset and 20% as the validation dataset. At last, three groups of data will be fed into the NN separately to evaluate their performance.

Data preprocessing is conducted with the smoothed method to reduce the random variation representing random measurement errors, which makes it harder to train a neural network model. Also, the MDS problem in this paper means that we focus on the degradation cost for a relatively long period which is typically 24 hours. Comparing with the unsmoothed data and the smoothed data, we found for the same cycle, the total battery degradation is almost the same for a 24 hour time period. However, the training accuracy of the NNBD model can be increased by smoothing out the input data while the prediction error does not increase with the smoothed data. The reason is that the unsmoothed data is quite unstable in some short period as shown in the Fig. 1, which increases the difficulty of machine learning model training and decrease the training accuracy. However, if we look at the degradation for a long time period, the smoothed data and the unsmoothed data have a similar degradation value as shown in Table II. The cumulative degradation numbers in Table II are calculated based on the battery aging test with 60% SOC, 20% DOD, 0.75 C rate and 32℃ ambient temperature. There are five cycle intervals analyzed in Table II; from the results, we can observe that the total degradation value for the certain interval between the raw data and the pre-processed data are very similar. Also, when the length of the interval increases, the difference between the raw and the pre-processed data decreases. This ensures that the data pre-processing method improves the training accuracy without affecting the cumulative degradation value.

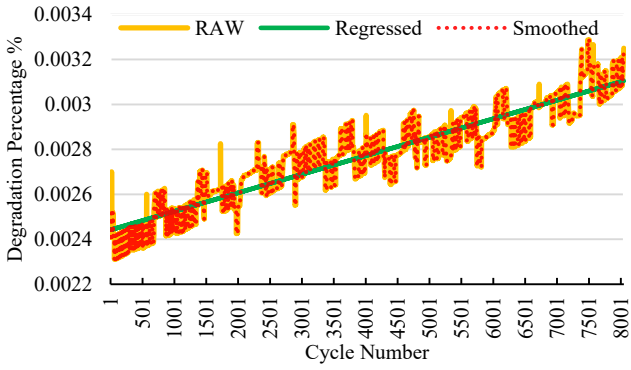

Fig. 1. Battery degradation data under different data processing methods.

Table II The cumulative degradation comparison over raw data and pre-processed data

| Cycle Range | Raw data degradation | Pre-processed degradation | Difference |
|---|---|---|---|
| 0-500 | 1.213% | 1.231% | 1.5% |
| 0-1000 | 2.454% | 2.483% | 1.18% |
| 0-1500 | 3.711% | 3.752% | 1.10% |
| 0-2000 | 5.007% | 5.048% | 0.82% |
| 0-2500 | 6.347% | 6.362% | 0.24% |

### C. The Proposed NNBD Model

A fully connected neural network is constructed to model the battery degradation. Five aging factors (ambient temperature, C rate, SOC, DOD and SOH) form a five-element input vector for the neural network. Each input vector corresponds to a single output value which is the amount of battery degradation in percentage respect to the SOH level for the same cycle. A dynamic learning rate scheme is used in the training process to improve the training result. The learning rate will decrease automatically after a certain number of training epochs. The structure of the trained neural network is shown in Fig. 2 plotted by NN-SVG [38]. It has an input layer with 5 neurons, first hidden layer with 20 neurons, second hidden layer with 10 neurons and an output layer with 1 neuron. The activation function for the hidden layers is "relu" and "linear" for the output layer.

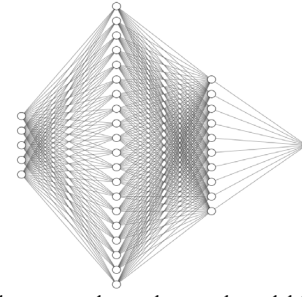

Fig. 2 Structure of the proposed neural network model for battery degradation.

Mini batch gradient descent strategy is applied to train the neural network. Different batch sizes are tested to achieve the best training results. As defined in (18), mean squared error (MSE) represents the average of the square of the difference between the actual and predicted values over all training data points. MSE is used as the criterion, a loss function, to train the neural network and measure the neural network quality.

$$MSE = \frac{1}{n}\sum_{i=1}^{n}(y_i - \widetilde{y}_i)^2 \quad (18)$$

### D. Battery Degradation Calculation

From the BESS operation profile, the SOC level is required as the input of the proposed NNBD model. The absolute value of the difference in the SOC levels between time intervals $t$ and $t-1$ will be the DOD level as shown in (19). C rate is calculated by (20). It is assumed that the battery SOH level is available prior to the microgrid day ahead scheduling. The input vector can be formed as shown in (21) and then fed into the trained NNBD model to obtain the total battery degradation over the MDS time horizon in (22).

$$DOD_t = |SOC_t - SOC_{t-1}| \quad (19)$$
$$C_t^{Rate} = DOD_t/\Delta T \quad (20)$$
$$\overline{x}_t = (T, C, SOC, DOD, SOH) \quad (21)$$
$$BD = \sum_{t \in S_T} f^{NN}(\overline{x}_t)SOH \quad (22)$$

### E. Cycle Based Battery Usage Processing Method

The equation (22) for battery degradation calculation mentioned in the previous section covers all the time periods for the microgrid day ahead scheduling. However, the BESS is not scheduled in fixed cycles with constant charge/discharge rates per MDS, leading to inconsistency with the fixed-cycle based data that are used for training NNBD model which is the gap mentioned in the introduction section. Furthermore, the idle status is also considered in (22) but it is out sampled for the NNBD model. In other words, equation (22) considers each time interval as a BESS cycle and feeds the associated data directly into the NNBD model, which may not accurately predict the BESS degradation. Therefore, the proposed cycle based battery usage processing method is developed to address the inconsistency and accurately apply the proposed NNBD model in the MDS problem.

In the CBUP method, instead of considering each time interval as a cycle, the operating time intervals of BESS scheduling are combined and averaged into different cycles. For any continuous time intervals, if the operation status (charging or discharging) does not change, they will be aggregated as a single charging or discharging cycle. For the aggregated cycle, the charging or discharging power will be the average of the aggregated time periods. The SOC will be the initial SOC val-ue of the first time period in the aggregated time periods and the DOD will be the absolute value of the SOC difference between the start time period and the end time period respectively. Thus, equation (22) is replaced by (23) while $\overline{x}_c$ represents the input vector for the aggregated cycles and $AC$ represents the set of feasible aggregated cycles.

$$BD = \sum_{c \in AC} f^{NN}(\overline{x}_c) SOH \quad (23)$$

*F. Incorporating NNBD into Microgrid Scheduling*

When considering battery degradation in microgrid day-ahead scheduling, the objective function needs to be updated by including the associated equivalent battery degradation cost. The updated objective function is shown in (24),

$$f = f^{MG} + f^{BESS} \quad (24)$$

where $f^{MG}$ is defined in equation (1); and $f^{BESS}$ denotes the battery degradation cost that can be estimated by the proposed NNBD model. $f^{BESS}$ can be calculated as follows,

$$f^{BESS} = \frac{c_{BESS}^{Capital} - c_{BESS}^{SV}}{1 - SOH_{EOL}} BD \quad (25)$$

where $c_{BESS}^{Capital}$ represents the capital investment cost of BESS; $c_{BESS}^{SV}$ denotes the salvage value at the end of life; $SOH_{EOL}$ represents state of health value that the BESS is considered as the end of life; $BD$ represents the percentage battery degradation calculated by (23).

Thus, the proposed battery degradation based MDS model can be represented by (2)-(15), (19)-(21), and (23)-(25).

## IV. THE PROPOSED NNODH FOR MICROGRID SCHEDULING

The proposed battery degradation based microgrid day-ahead scheduling model, which is presented in the above section, would be very hard to solve directly since the proposed NNBD model that is highly non-linear and non-convex is now a part of the proposed BDMDS model. To address this issue, a novel algorithm, neural network and optimization decoupled heuristic, is proposed in this paper to decouple the complex BDMDS model and solve it in the following five steps in an iterative manner.

- Step A is to solve the microgrid day-ahead scheduling with additional constraints by limiting BESS usage that is generated from Step E. Note that in the first iteration, there is no extra limit on BESS usage.
- Step B obtains the scheduled BESS operating profile that is then processed with the proposed CBUP method.
- Step C estimates BESS degradation with the proposed NNBD model that is trained beforehand and calculates the associated equivalent battery degradation cost.
- Step D determines whether the solutions meet the designed stopping criteria: stop the iteration process and report the solution if yes; otherwise, go to Step E.
- Step E updates the boundaries for the BESS related operating constraints from section IV.B which can limit the BESS operations to reduce the degradation cost; and the associated constraints will be sent to Step A to be included in MDS for the next iteration.

This iterative procedure is illustrated in Fig. 3.

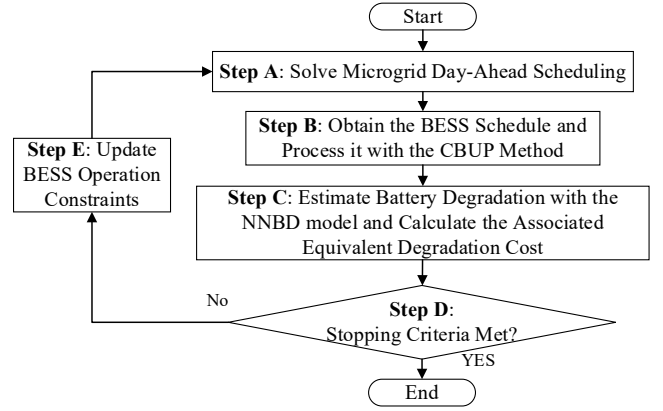

Fig. 3. Flowchart of the proposed NNODH algorithm.

*A. Microgrid Scheduling*

The traditional microgrid day-ahead scheduling that introduced in Section II is conducted to obtain an initial solution which does not consider the battery degradation. It is solved only in the first iteration. When considering battery usage limiting constraints that will be presented in Section IV.D, Step A will solve the CMDS instead of the traditional MDS. To sum up, traditional MDS is solved in the first iteration while CMDS is solved in all subsequent iterations.

*B. Operation Limits of BESS*

The goal of Step E is to generate new constraints that can limit or change the battery output to reduce the total battery degradation. The generated constraints will further tighten the range that BESS can operate after each iteration. Three types of extra constraints are proposed to limit battery output. Constraint (26) named as battery consumption limit (BCL) is to reduce the sum of battery output power over 24 hours by forcing it to be less than the previous iteration. Constraint (27) only limits the sum of output power in three time intervals that have the highest battery charging or discharging power, named as precise battery consumption limit (PBCL). Constraints (28)-(29) are designed to limit the maximum charging/discharging power that is named as battery C rate limit (BRL). The proposed constraints will be generated and updated in each iteration by using the scheduled battery operational profile from previous iteration and the proposed battery operation restriction factor (BORF) $\alpha$ which is a preset parameter. Note that the value of BORF can determine the limits in BCL, PBCL, and BRL for each iteration.

$$\sum_{t \in T}(P_{Char}^{t,i} + P_{Disc}^{t,i}) \leq (1-\alpha) * P_{BatteryTotal}^{Pervious\_SCUC} \quad (26)$$

$$\sum_{t \in Top\ 3} P_{Disc}^{t,i} + P_{Disc}^{t,} \leq (1-\alpha) * P_{BatteryTotal}^{Pervious\_SCUC} \quad (27)$$

$$P_{Char}^{t,i} \leq P_{Si}^{Max}(1-\alpha)^{iteration-1} \quad (28)$$

$$P_{Disc}^{t,i} \leq P_{Si}^{Max}(1-\alpha)^{iteration-1} \quad (29)$$

Depending on which battery usage limiting constraint is used in CMDS, there are four variations of the proposed NNODH algorithm. They are all conducted to compare the performance of the proposed constraints. The proposed four variations are presented in Table III. The CMDS model in this table is solved instead of the traditional MDS model starting from the second iteration of the proposed NNODH algorithm.

Table III The proposed four NNODH strategies

| Strategies | CMDS Model for Step A |
|---|---|
| NNODH-BCL | (1)-(15), (26) |
| NNODH-PBCL | (1)-(15), (27) |
| NNODH-BRL | (1)-(15), (28)-(29) |
| NNODH-ALL | (1)-(15), (26)-(29) |

### C. Iterations and Stopping Criteria

Fig. 4 illustrates the iteration process of determining the optimal solution for the battery degradation based MDS using the proposed NNODH algorithm. As is discussed in Section IV.A, the first iteration only considers the traditional microgrid day-ahead scheduling model that ignores the battery degradation. The NNODH algorithm can assure that the solution is guaranteed to reach optimal. The first iteration will provide the optimal solution when the battery degradation is not considered. The BESS operation will be compressed or limited to reduce the battery degradation in the rest of the iterations. During the iterations, the battery degradation will decrease gradually while the operation cost will increase gradually. Thus, the vertex point of the total cost curve is the optimal solution. In Fig. 4, the red star denotes the global optimal solution for the battery degradation based MDS that considers the proposed NNBD model; and the black star represents the solution for MDS/CMDS problem in the current iteration while the grey star denotes the solution for the MDS/CMDS problem in the previous iteration. The MDS/CMDS solutions are approximations to the global optimal solution for the battery degradation based MDS; they are referred to as pseudo solutions (feasible but may not be global optimal) for the battery degradation based MDS. Graph 1 in Fig. 4 represents the first iteration. The BESS usage limit constraints that generated in each iteration will reduce the feasible region for the MDS/CMDS problem in the next iteration. Graphs 2-4 show the feasible region shrinks after each iteration and the black star (pseudo solution) is moving towards the red star (global optimal solution for BDMDS). Graph 5 shows the red star and black star overlap which indicates that the black star here is the optimal solution for the BDMDS problem. If the iteration continues, the feasible solution area of BDMDS will continue to reduce such that the optimal solution might be cut out. Since the proposed NNODH algorithm is an iterative method, the $\alpha$ values (BORF) in (26)-(29) would change over iterations and may result into different BDMDS solutions, among which the best solution should deviate slightly away from the global optimal solution. To ensure the best pseudo solution is close enough to the global solution with least total cost, small $\alpha$ value is preferred. However, it may take more iterations for a smaller $\alpha$ value to find the optimal solution.

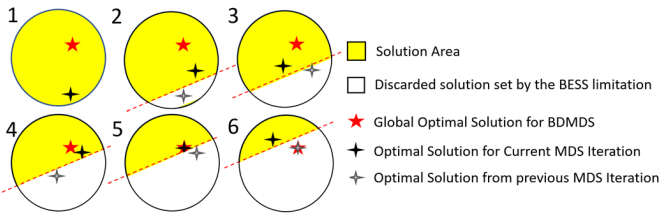

Fig. 4. Illustration of the proposed NNODH algorithm.

### D. Evaluation Metrics

Three metrics are designed to evaluate the performance of the proposed model: degradation cost reduction (DCR), total cost reduction (TCR) and operational cost increment (OCI). They are defined in (30), (31), and (32) respectively. The degradation cost reduction illustrates how much the battery degradation cost can be reduced as compared to the maximum value when battery degradation is not considered in traditional MDS. The TCR represents how much the total cost can be reduced by the proposed model while the OCI shows how much the operation cost would increase when considering the battery degradation model. Notations $BDC^{Max}, TC^{Max}, OC^{Min}$ represent the maximum battery degradation cost, maximum total cost, and the minimum operation cost respectively. These metrics are obtained by solving the traditional MDS problem in the first iteration when battery degradation cost is not considered.

$$DCR = \frac{BDC^{Max} - BDC}{BDC^{Max}} * 100\% \tag{30}$$

$$TCR = \frac{TC^{Max} - TC}{TC^{Max}} * 100\% \tag{31}$$

$$OCI = \frac{OC - OC^{Min}}{OC^{Min}} * 100\% \tag{32}$$

When we consider these proposed metrics to evaluate the proposed algorithm, the TCR will be the first choice since the objective of the algorithm is to reduce the total cost. The algorithm with a higher TCR is always preferred in the analysis. In the meanwhile, the DCR and OCI are designed to gauge the degradation cost and operation cost. We prefer a higher number of the DCR and a lower number of the OCI. However, DCR and OCI are not as important as the TCR here.

A stopping criterion is proposed as part of the NNODH algorithm. It is used to terminate the iterations and then the best solution can be reported. The designed stopping criteria is defined as follows: for any past 10 iterations, if the $TC^n - TC^{n-1}$ is less than 0 for the first five iterations but greater than 0 for the last five iterations, then the iteration will stop and the best solution with lowest total cost for BDMDS will be reported. Here, $TC^n$ and $TC^{n-1}$ denotes the total costs in $n^{th}$ and $(n-1)^{th}$ iterations respectively. The total cost here represents the lowest total cost rather than the lowest battery degradation cost. The optimal solution may be affected by the battery size as well as the unit price.

### E. Benchmark Models of MDS

Benchmark models are designed to evaluate and demonstrate the effectiveness and performance of the proposed NNODH algorithm. Three benchmark models are presented below and summarized in Table IV:

1. Traditional MDS Model: The basic benchmark model is set with no BESS degradation cost and the MDS will maximize the usage of BESS to minimize the operation cost of the microgrid.
2. Cycle Limit Model: This model is widely adopted in the industry. It is designed to limit the charging and discharging cycles to decrease the degradation of the BESS. This model is formulated with the following constraints where the $V_{Char/Disc}^{t,i}$ is a binary variable that represents the status change of charging or discharging operation.

$$V_{Char/Disc}^{t,i} \leq U_{Char/Disc}^{t,i} + U_{Char/Disc}^{t-1,i}, \forall i, t \tag{33}$$

$$V_{Char/Disc}^{t,i} \geq U_{Char/Disc}^{t,i} - U_{Char/Disc}^{t-1,i}, \forall i, t \tag{34}$$

$$V_{Char/Disc}^{t,i} \geq U_{Char/Disc}^{t-1,i} - U_{Char/Disc}^{t,i}, \forall i, t \tag{35}$$

$$V_{Char/Disc}^{t,i} \leq 2 - U_{Char/Disc}^{t,i} - U_{Char/Disc}^{t-1,i}, \forall i, t \quad (36)$$
$$V_{c/disc}^{1} = 0 \quad (37)$$
$$\sum_t V_{Char/Disc}^{t,i} \leq 2 \quad (38)$$

3. Linear Battery Degradation Cost (BDC) Model: The third benchmark model is based on constant battery degradation cost parameters. It is assumed that the battery degradation cost is linear to the energy consumption of the BESS as shown in (39) where the $c_{BESS}$ is a fixed rate that represents the battery degradation cost per unit usage of BESS.

$$f^{BESS} = c_{BESS} \sum_{t,i} P_{Char/Disc}^{t,i} * length(t) \quad (39)$$

Table IV Benchmark models

| MDS Models | Formulations |
|---|---|
| Traditional MDS | (1)-(15) |
| Cycle Limit | (1)-(15), (33-38) |
| Linear BDC | (1)-(15), (39) |

## V. CASE STUDIES

### A. Microgrid Testbed

A typical grid-connected microgrid with renewable energy sources is created in this paper as a testbed to examine the performance of the proposed NNBD model, BDMDS model and NNODH algorithm. This testbed includes one 180kW diesel generator (DG), five 200kW wind turbines (WT), 300 residential houses that contains solar panel (5kW capacity per house), and a 300kWh lithium-ion based BESS with a charging/discharging efficiency of 90%. The load data representing 1000 residential houses. The ambient temperature and available solar power for a time period of 24 hours are obtained from the Pecan Street Dataport [39]. The wholesale electricity price is obtained from ECROT [40]. The price of the electricity sold to the main grid is set to 80% of the purchase price. Sensitivity analysis is conducted with different RES penetration levels and different BESS sizes. The computer with Intel® Xeon(R) W-2295 CPU @ 3.0 GHz, 256 GB of RAM, and Nvidia Quadro RTX 8000 (48GB GPU) was utilized to conduct the numerical simulations including the training of battery degradation model and the optimization of microgrid day-ahead scheduling. The microgrid resource scheduling problem studied in this work covers a total time horizon of 24 hours, which is solved by the Pyomo package [41] with the Gurobi solver [42].

### B. Training Results of NNBD

The training results of NNBD and hyperparameters are presented and analyzed in this section. Mini-batch technology is used for the NN training. The optimal batch size may vary for different input data. Different training batch sizes are tested to determine the optimal training batch-size and the test results are presented in Table V. It can be observed that the batch size of 256 can achieve the highest accuracy while requiring less epochs to complete the training process. The validation accuracy can reach up to 94.5% while it only requires 50 epochs to reach a steady accuracy. The error tolerance is set to 15% when calculating prediction accuracy in Table V and Table VI. Larger or smaller batch size will either reduce the accuracy or increase the training epochs. We found that large batch size can help smooth the oscillation of the training accuracy curve. The training accuracies with batch sizes of 128 and 256 are almost equally the best. However, the training accuracy curve with a batch size of 256 is much smoother than the other one. Moreover, we observed that if the input data are shuffled, then the neural network cannot obtain good results. This may be due to the characteristic of the input data: the battery degradation data are time-series for each battery aging test. The results with different data pre-processing methods are compared in Table VI. Note that the test results in Table VI are obtained with the batch-size of 256 for all the trainings. The regressed data pre-processing method has the highest accuracy and efficiency. Based on the results from Table V and Table VI, the regressed method performs the best and is applied for all subsequent simulations.

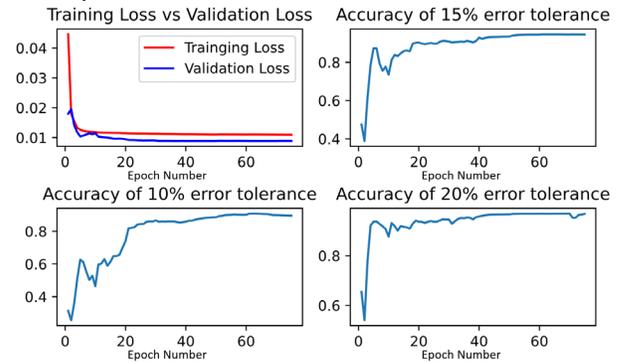

Fig. 5. DNN training and validation results.

Table V Sensitivity test of batch-sizes

| Batch-Size | 16 | 32 | 64 | 128 | 256 | 512 | 1024 | 2048 |
|---|---|---|---|---|---|---|---|---|
| Number of Epochs | 20 | 20 | 85 | 110 | 50 | 50 | 75 | 75 |
| Accuracy | 40% | 50% | 90.5% | 94.0% | 94.5% | 92.5% | 88% | 67% |

Table VI Training results with different data pre-processing methods

| Data Pre-processing Methods | Accuracy | Number of Epochs |
|---|---|---|
| Raw | 78% | 150 |
| Smoothed | 82% | 100 |
| Regressed | 94.5% | 50 |

The best training results are shown in Fig. 5 that illustrates the training loss versus validation loss and the accuracy curves under different error thresholds. The accuracy is around 60% under a 5% error tolerance, 80% under a 10% error tolerance and 94.5% under a 15% error tolerance. The accuracy with 20% error tolerance is 95.5% which is only 1% higher than the 15% error tolerance. Thus, we choose 15% as the error tolerance level when calculating accuracy in all subsequent results. After 50 epochs, the training loss and the validation loss stop decreasing and the training accuracy becomes stable as well. The training stops at $65^{th}$ epoch to avoid overfitting and oscillation.

### C. Results of NNODH algorithm

In this section, the BDMDS results obtained with the NNODH algorithm are presented. The load profile of the test bed is shown in Fig. 6. Table VII presents the results for different strategies of the proposed NNODH algorithm presented in Section IV. The BCL, PBCL, BRL and ALL CMDS models are implemented and tested separately. The combination of all three types of constraints as the added constraints to MDS in the next iteration, marked as ALL in Table VII, is tested as well. In Table VII, for all the proposed models, the initial iteration does not have any limits of the BESS operation. This also leads to the solution of the first iteration having the highest battery degradation cost. Similarly, the maximum total cost for different models is the same. For metric DCR, the ALL option performs the best among all strategies, and it decreases the battery degradation cost by 79.27%. BCL performs the best in terms of metric TCR. The increased operation cost is similar between BCL and PBCL, which is less than BRL and ALL. Overall, we prefer the NNODH-BCL strategy due to its best performance to decrease the total cost and the fast solving time.

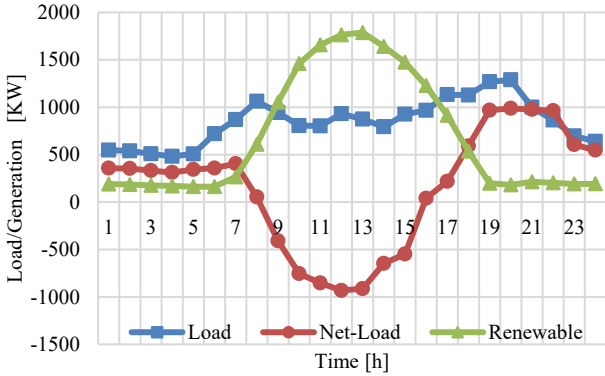

Fig. 6. Load profile of the microgrid testbed.

Table VII Results for proposed strategies

| NNODH | BCL | PBCL | BRL | ALL |
|---|---|---|---|---|
| Total Cost ($) | 494.36 | 503.49 | 502.50 | 500.32 |
| Degradation Cost ($) | 10.74 | 19.50 | 11.99 | 10.39 |
| TCR | 5.82% | 4.08% | 4.27% | 4.69% |
| DCR | 78.57% | 60.23% | 76.08% | 79.27% |
| OCI | 1.83% | 1.81% | 3.20% | 3.09% |
| Solving time (s) | 8.68 | 12.82 | 10.00 | 8.48 |
| Iteration Numbers | 37 | 56 | 46 | 37 |

Fig. 7 shows the results for NNODH-BCL including the degradation cost, operation cost, battery degradation in percent, and total cost. These results are based on the battery unit price of 400 $/kWh and the value of BORF is set to 0.03. From Fig. 7, we can observe that the valley of the total cost curve is at the 37$^{th}$ iteration with $494.36 including the equivalent battery degradation cost of $10.74 and the microgrid operation cost of $483.62. The total cost is reduced by 5.82% compared to the traditional MDS model that does not consider the battery degradation cost, which proves that the proposed NNODH algorithm can reduce the total cost significantly. Note that the results shown in Fig. 7 did not implement the stopping criteria to show how the system behaves when the battery's usage is further limited until idle. If not, the iteration will stop at the 42$^{nd}$ iteration since the minimum total cost can be found at the 37$^{th}$ iteration. After the 37$^{th}$ iteration, we can observe that even though the battery degradation cost keeps decreasing, the total cost starts to increase due to the increased slope of the operation cost.

The scheduled BESS operation for different models are shown in Fig. 8. Positive output means the BESS is in discharging mode and negative output means it is in charging mode. The scheduled BESS operations for the Traditional MDS model, Cycle Limit model, Linear BDC model and the proposed BDMDS model are all shown in Fig. 8. It can be observed for the traditional MDS that does not consider the battery degradation, BESS operates at a wider output range from -150 kW to 150 kW in seven different time intervals. When battery degradation is considered, BESS is scheduled to charge and discharge in a narrow range and in less active time intervals. The BESS operation patterns for different models match in most of the time intervals, which proves the effectiveness of those four models. For the proposed BDMDS, after applying the NNODH algorithm to solve it, the BESS operates only in three different time intervals. Total exchanged energy is limited to reduce the battery degradation which meets the designed purpose of constraint BCL. The total usage of BESS for Traditional MDS model is 920 kWh while it is 325 kWh for BDMDS, 601 kWh for Cycle Limit model and 473 kWh for Linear BDC model.

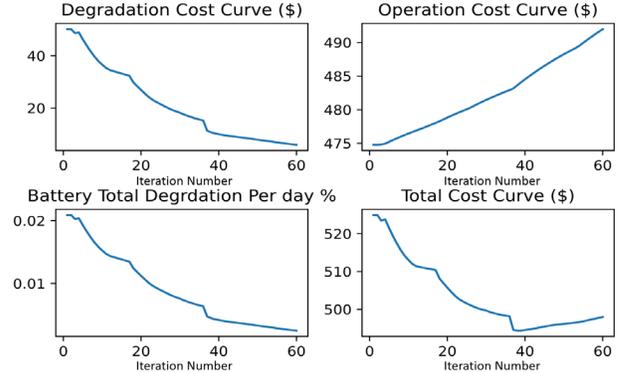

Fig. 7. BDMDS Results of the NNODH-BCL method.

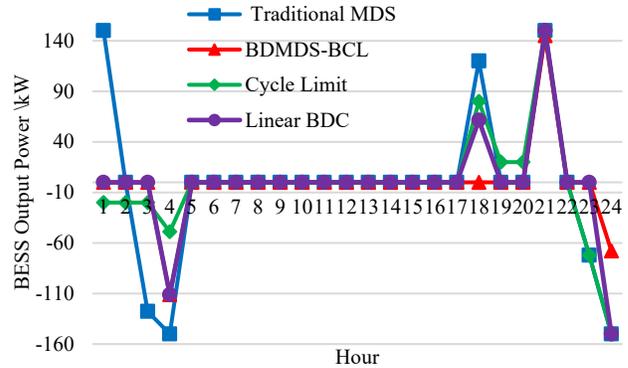

Fig. 8. BESS scheduled operations comparison.

Table VIII presents the results of the CBUP method applied on the solutions obtained from BDMDS model and traditional MDS model. It can be clearly observed that the battery degradation prediction is pretty close to the true battery degradation in a small percent of error. Without the CBUP method, the error of BD prediction from NNBD model is significantly higher. Thus, the results proves the effectiveness of the proposed CBUP method. From Table IX, we can observe that the proposed BDMDS model has the lowest daily degradation and the lowest annual degradation cost among all the models. Moreover, the BDMDS model is outstanding in annual cost saving. Also, if we consider the load profile is the same for each day, then the proposed BDMDS model can significantly extend the lifetime of the BESS than all the benchmark models. Note that the expect end of life is set to 70% of SOH value.

Table X shows the results of the sensitivity tests with different BORF values. The results show that larger BORF can achieve the lowest total cost in less iterations and less solving time. As a tradeoff, the optimal total cost of larger BORF will be slightly higher. Smaller BORF will lead to a smaller area cut from the feasible solution area in each iteration, which requires more iterations to converge to the optimal solution. However, the optimal solutions for the different BORFs do not have a significant difference. Thus, a higher value of BORF such as 0.05 is preferred due to the high computing efficiency.

Table VIII Battery degradation prediction with different methods

| Model | True BD | Without CBUP | | With CBUP | |
|---|---|---|---|---|---|
| | | Degradation | Error | Degradation | Error |
| BDMDS | 0.00468% | 0.037% | 690% | 0.0044% | 5.9% |
| MDS | 0.0108% | 0.066% | 511% | 0.0120% | 9.7% |

Table IX Model comparison

| Model | Daily BESS Degradation | Annual Degradation Cost ($) | Annual Cost Saving ($) | Expect Lifetime (years) |
|---|---|---|---|---|
| MDS | 0.02% | 18,301.1 | N/A | 4.1 |
| Cycle Limit | 0.012% | 12,540.8 | 6,205 | 6.8 |
| Linear BDC | 0.01% | 8,832.5 | 6,935 | 8.2 |
| BDMDS | 0.0045% | 3,920.1 | 11,151 | 18.3 |

Table X Results of sensitivity analysis with different BORF values

| BORF | Number of Iterations | Optimal Total Cost ($) | Time (s) |
|---|---|---|---|
| 0.01 | 113 | 494.35 | 25.21 |
| 0.02 | 57 | 494.35 | 12.51 |
| 0.03 | 38 | 494.36 | 8.42 |
| 0.05 | 22 | 494.36 | 5.20 |
| 0.1 | 11 | 494.38 | 2.85 |
| 0.2 | 5 | 494.40 | 1.41 |

### D. Sensitivity Analysis of RES Penetration Levels

Table XI presents the TCR and OCI results under different RES penetration levels for the microgrid. The penetration level is defined as the ratio of the average renewable generation to the average load in a typical day. The proposed microgrid testbed is set to an 80% RES penetration level. From Table XI, we can observe that with the increase of RES penetration level, the value of TCR will increase and the value of OCI will decrease. This is because that in a system with higher RES penetration, BESS is required to charge/discharge more frequently to mitigate the uncertainty resulted by the RES generation, which leads to a higher battery degradation.

Table XI Results of different RES penetration levels

| | RES Penetration Level | | | |
|---|---|---|---|---|
| | 20% | 40% | 60% | 80% |
| TCR | 1.60% | 2.1% | 3.12% | 5.82% |
| OCI | 0.47% | 0.63% | 0.94% | 1.83% |

### E. Sensitivity Analysis of BESS Unit Prices and Sizes

In this section, the sensitivity analysis on different BESS unit prices and sizes is conducted. Fig. 9 shows that the total cost reduction in percentage. It can be clearly observed that for the same size of BESS, higher unit price corresponds to a higher total cost reduction in percent. For the same unit price of BESS, higher BESS size tends to achieve a lower TCR. Table XII shows that for the same unit price except $200/kWh, higher BESS size has a lower DCR. This may be because that for the same battery output power, bigger battery size will likely have a lower C rate which is one of the main contributing factors of battery degradation. Also, for the same size BESS, higher unit price lead to a higher DCR. Thus, we can conclude that the BESS price of per unit capacity is the major factor affecting how much the proposed model reduces the degradation cost. The BESS size is the major factor affecting how much the proposed model reduces the battery degradation. The result of the sensitivity analysis in this section may help determine the size of the BESS for microgrid planning. In the meanwhile, as the unit price of BESS keeps decreasing, the battery degradation cost will be lower and account for smaller percentage respect to the total cost in the future.

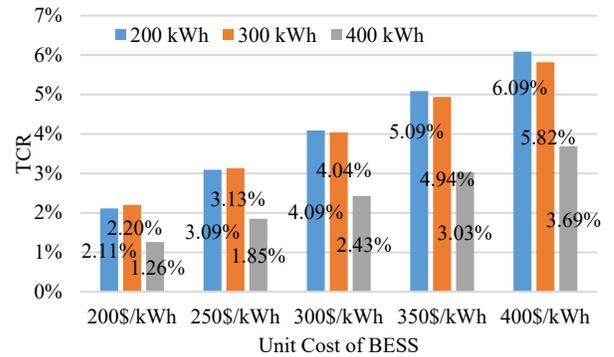

Fig. 9. Results of TCR sensitivity tests.

Table XII DCR of different BESS sizes and unit prices

| | Unit Price ($/kWh) | | | | |
|---|---|---|---|---|---|
| Size (kWh) | 200 | 250 | 300 | 350 | 400 |
| 200 | 75.15% | 82.16% | 82.96% | 85.18% | 88.58% |
| 300 | 77.29% | 77.29% | 78.57% | 78.57% | 78.57% |
| 400 | 52.95% | 54.48% | 54% | 71.90% | 72.68% |

## VI. CONCLUSIONS

In this paper, a neural network based battery degradation model is proposed to predict the BESS degradation value for each scheduling period. A cycle based battery usage processing method is designed to accurately apply the proposed NNBD model with the microgrid day-ahead scheduling model. The NNODH algorithm is proposed to decouple the battery degradation based microgrid day-ahead scheduling problem that is hard to solve directly due to the highly non-linear characteristic of the proposed NNBD model. The proposed NNODH algorithm can solve the MDS optimization problem and calculate the battery degradation cost iteratively and effectively find the optimal solution with the lowest total cost. The CMDS problem is created with the options of using different proposed constraints to limit the BESS usage and obtain a lower battery degradation cost. An RES-enriched microgrid is used to evaluate the performance of the proposed BDMDS model and the proposed NNODH algorithm.

The test results demonstrate that the battery degradation can be accurately predicted (5.9% error) by the proposed NNBD model with the adjusted inputs obtained by the proposed CBUP method. The proposed NNODH algorithm can obtain the optimal solution efficiently. Compared with the traditional

MDS models, the total cost can be reduced significantly by 5.82% with the proposed BDMDS model. Also, the proposed model can reduce the daily BESS degradation significantly from 0.02% to 0.0045%. The annual degradation cost is reduced by 78.6% with the proposed model. Moreover, the expected lifetime is extended from 4.1 years to 18.3 years with the proposed model. The NNODH-BCL performs the best among the proposed four strategies in this case. Sensitivity tests demonstrate the performance of the proposed NNODH algorithm under different BESS sizes and unit prices. Overall, this work demonstrates the effectiveness of the proposed BDMDS model for reducing battery degradation cost and total cost, and the capability of the proposed NNODH algorithm for efficiently solving BDMDS that is a deep neural network embedded optimization problem.

## VII. Future Work

The uncertainty of the renewable generation can be considered and addressed with advanced models such as a stochastic and robust optimization model, which is a potential extension of this work. Moreover, battery aging test data are often very limited, not supporting deep learning studies that require large amounts of data. One solution we plan to develop to address this issue is to leverage transfer learning [43] with the trained NNBD model in this paper to obtain different NN models for other types of batteries. This would enable us to train battery degradation models for different types of batteries with much less data.